\documentclass{aa}

\newcommand{\MSun}  {\mbox{${\rm M}_{\sun}$}}
\newcommand{\MJup}  {\mbox{${\rm M}_{\rm Jup}$}}

\newcommand{\RHill} {\mbox{${\rm R}_{\rm Hill}$}}

\newcommand{\AU}    {\mbox{${\rm AU}$}}
\newcommand{\kms}   {\mbox{${\rm km\,s^{-1}}$}}

\usepackage[varg]{txfonts}
\usepackage{amsmath}
\usepackage{amssymb}
\usepackage{url}
\usepackage{stmaryrd}
\usepackage{graphicx}
\usepackage{natbib}
\usepackage{bm}
\usepackage[T1]{fontenc}
\usepackage{ae,aecompl}
\usepackage{hyperref}
\usepackage[all]{hypcap}
\usepackage{color}

\definecolor{darkblue}{rgb}{0.0, 0.0, 0.5}
\hypersetup{colorlinks=true,citecolor=darkblue,linkcolor=darkblue}

\title{Constraints on the size and dynamics of the J1407b ring system}
\titlerunning{Constraints on the size and dynamics of the J1407b ring system}

\author{
  Steven~Rieder\inst{\ref{inst:aics},\ref{inst:strw}}\thanks{steven@rieder.nl}
  \and Matthew~A.~Kenworthy\inst{\ref{inst:strw}}
}
\authorrunning{S.~Rieder \and M.A.~Kenworthy}

\institute{
  RIKEN Advanced Institute for Computational Science, 7-1-26
  Minatojima-minami-machi, Chuo-ku, Kobe, Hyogo 650-0047,
  Japan\label{inst:aics}
  \and Sterrewacht Leiden, Leiden University, P.O. Box 9513, 2300 RA
  Leiden, The Netherlands \label{inst:strw}
}

\begin{document}

\abstract {
  J1407 (\object{1SWASP J140747.93-394542.6} in full) is a young star in the
  Scorpius-Centaurus OB association that underwent a series of complex
  eclipses over 56 days in 2007.
  To explain these, it was hypothesised that a secondary substellar
  companion, J1407b, has a giant ring system filling a large fraction
  of the Hill sphere, causing the eclipses.
  Observations have not successfully detected J1407b, but do rule out
  circular orbits for the companion around the primary star.
} {
  We test to what degree the ring model of J1407b could survive in an
  eccentric orbit required to fit the observations.
} {
  We run $N$-body simulations under the AMUSE framework to test the
  stability of Hill radius-filling systems where the companion is on
  an eccentric orbit.
} {
  We strongly rule out prograde ring systems and find that a secondary
  of $60$ to $100 \MJup$~with an 11 year orbital period and retrograde
  orbiting material can survive for at least $10^4$ orbits and produce
  eclipses with similar durations as the observed one.
} 


\maketitle

\section{Introduction}
\label{sec:intro}

Giant planet formation consists of the transfer of material from the
circumstellar environment to the circumplanetary environment.
The change in angular momentum of the circumstellar material results
in the formation of a disk of gas and dust surrounding the
protoplanet \citep{Ward10,Alibert05}.
In our Solar system, the primordial gas is no longer present, but
evidence of the circumplanetary disk exists in the form of coplanar
moons and rings \citep[e.g. see review by][]{Tiscareno13}.
All Solar system gas giant ring systems show structure. This structure
consists of gaps in the rings themselves and sudden changes of particle
density as a function of radius from the planet.

The K5 pre-MS 16 Myr-old star J1407 showed a complex series of
eclipses in 2007, lasting a total of 56 days, and a series of papers
investigating the J1407 system \citep{Mamajek12, vanWerkhoven14,
Kenworthy15, Kenworthy15b} conclude that there is a secondary
substellar companion (called J1407b) with a giant multi-ring system in
orbit around the primary star.
The ring system shows detailed structure down to the temporal
resolution set by the diameter of the primary star and their mutual
relative projected velocity.
A study of the stability of a Hill sphere-filling system on a circular
orbit has been explored in \citet{2016arXiv160502365Z}.
The derived diameter of the ring system combined with the
observational limits set on the companion J1407b \citep[as described
in][]{vanWerkhoven14,Kenworthy15} imply that J1407b is on an eccentric
orbit about J1407.

In this paper we investigate the effects of an elliptical orbit on the
stability of the ring system surrounding J1407b.
Our goal is to determine whether there are any bound orbital solutions
for the secondary companion that can explain both the derived relative
velocity and the duration of the eclipse seen towards J1407.

To this end, we construct a model containing J1407, its companion
J1407b and a co-planar disc around J1407b; based on the model in
\citet{Kenworthy15b}. In Section~\ref{sec:model}, we describe this
model and the parameters we investigate in this article. 
We run simulations of this model using {\tt
AMUSE}\footnote{\url{http://amusecode.org}} \citep{Pelupessy13,
PortegiesZwart13} with the {\tt
Rebound/WHFast}\footnote{\url{http://rebound.readthedocs.io}}
\citep{Rebound, WHFast} $N$-body integrator. We give the resulting
disc sizes and eclipse durations in Section~\ref{sec:size} and discuss
our results and the consequences in Section~\ref{sec:results}.

\section{Description of the J1407b ring models}
\label{sec:model}

To investigate how the extent of a ring system around J1407b would
change over time, we create a model consisting of J1407, companion
J1407b on an eccentric orbit around J1407 and a disc of $N$ particles
around J1407b, initially in either prograde or retrograde circular
orbits.  Since the height to diameter ratio of the J1407b exorings is
$< 0.01$ \citep{Mamajek12}, we limit our model to be co-planar with
the secondary's orbit. 

For the orbital parameters of J1407b, we choose values based on the
best fits found by \citet{Kenworthy15}. For all models, we choose a
period of $11$ years, resulting in a semi-major axis $a$ of $5.0\pm
0.1$ AU, and a mass $M$ of J1407 of $1.0\MSun$.  
We set the J1407b mass $m$ to values between $20$ and $100 \MJup$ in
steps of $20 \MJup$, as \citet[Figure~15]{Kenworthy15} constrained the
mass of J1407b for elliptic orbits to be up to 100\MJup.

We assume that the eclipse was observed at or very near its
pericentre, since the orbital velocity is highest at this point and
J1407b would likely have been detected if the long axis of the orbit
were not aligned to the line of sight \citep{Kenworthy15}.

\begin{figure}
\centering
\includegraphics[width=\hsize]{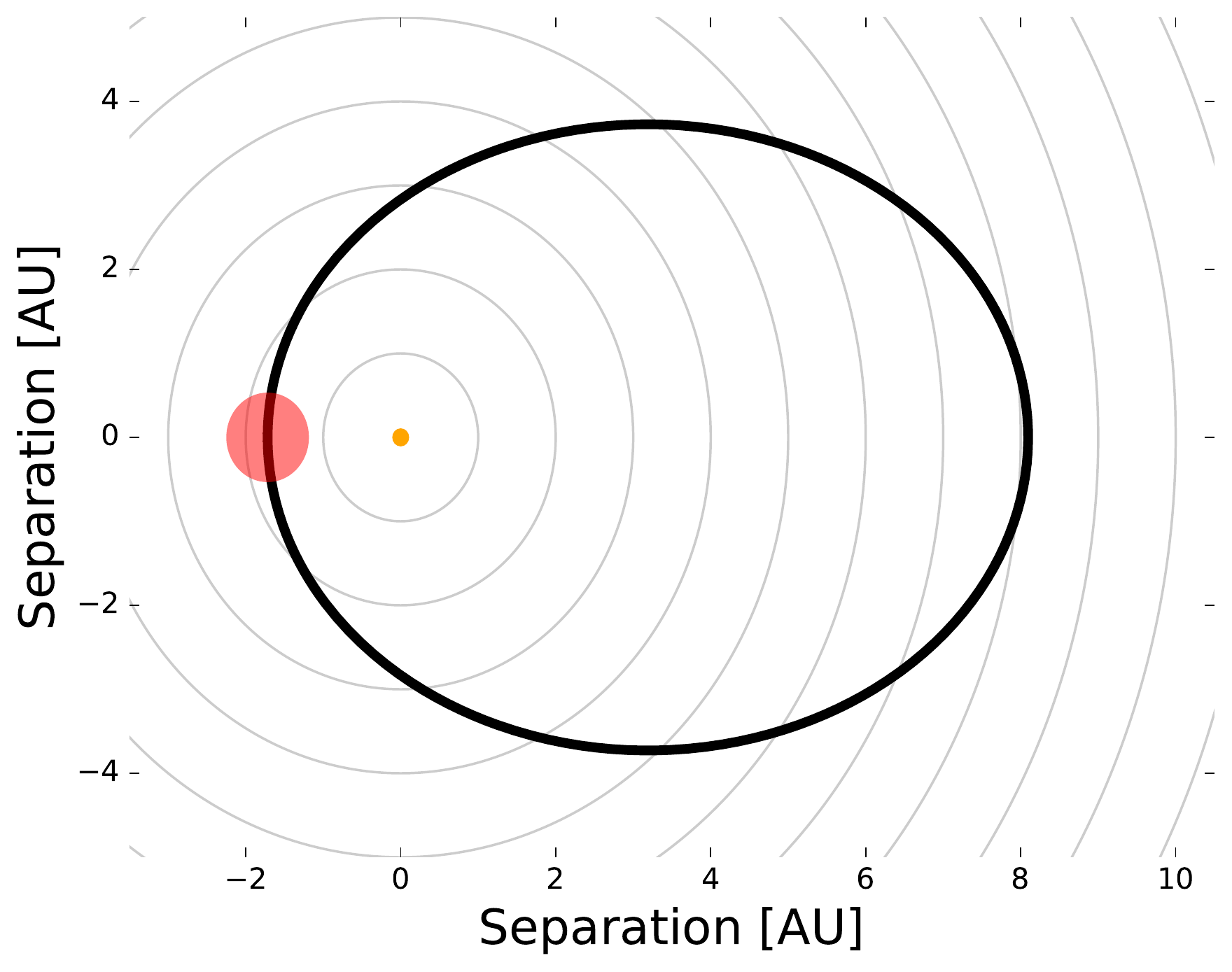}
  \caption{The orbit of J1407b model B80, with J1407b located at
    pericentre. The J1407b system (red) is shown to scale for the
    initial size of the model. The size of the star (orange) is
    exaggerated by a factor 20. Grey circles indicate the distance to
    the star in AU, while the black ellipse shows the orbit.
    \label{fig:orbit} }
\end{figure}

The pericentric velocity of the system is $32\pm2\kms$
\citep[Figure~11]{Kenworthy15}.  This velocity is inferred from the
stellar radius and the steepest gradient in the light curve, with the
second-steepest gradient resulting in a velocity of $18\kms$.  To
account for these velocities, while also investigating a slightly less
extreme case, we choose values between $0.6$ and $0.7$ for the
eccentricity $e$. This results in pericentric velocities between $27$
and $33~\kms$ and pericentric distances between $1.5$ and $2.0~\AU$. 
In Figure~\ref{fig:orbit} we show the orbit of model B80 as an
example. We summarise the selected values in
Table~\ref{tab:modelparam}.

\begin{table}
  \centering
  \caption{Orbital parameters of J1407b used for the different models.
    \label{tab:modelparam}}
  \begin{tabular}{l l l l l l l}
    Model &  e   & $v_{\rm peri}$ & $m$ \\
          &      & ($\kms$)       & ($\MJup$)\\
    \hline
    A     & 0.70 & $32.5\pm 0.4$    & 20,40,60,80,100 \\
    B     & 0.65 & $29.5\pm 0.4$    & 20,40,60,80,100 \\
    C     & 0.60 & $27.3\pm 0.3$    & 20,40,60,80,100 \\
  \end{tabular}
\end{table}

We initially place particles on equidistant rings, with equal distance
between particles in each ring. We choose the number of particles in
the $n$th ring to be $1 + 6 n$.  To prevent artificially created
rings, we then change the radial distance of each particle by a random
value drawn evenly between $\pm 0.5\times \Delta R$ (with $\Delta R$
the distance between rings).  We create 50 rings, scaling the system
so that the outer ring is initially at the Hill radius of the system
calculated at the companion's pericentre (Equation~\ref{eqn:Hill}
\citet{HamiltonBurns92}). This results in initial radii ranging from
$0.29\AU$ (for model C-20) to $0.66\AU$ (for model A-100).
\begin{equation} 
  \RHill = a (1 - e) \left( \frac{m}{3 M} \right)^{1/3}
  \label{eqn:Hill}
\end{equation}

Since the star's influence is smallest for the innermost orbits, we
limit the inner radius to $0.25\times \RHill$. This eliminates the
first 12 rings of particles, while a significant fraction of orbits
within the Hill radius is covered.  As the timestep of our simulation
scales with the orbital period of the particles, this also speeds up
our simulation. The total number of disc particles in each of our
models is $N=6992$.

In this article, we limit ourselves to investigating the extent to
which a disc can survive, while the creation and evolution of ring
structure is left to future study. Since the mass of the disc is
negligible compared to the secondary's mass, we ignore the internal
dynamics of the disc and make the disc particles massless, speeding up
the calculations. 
At the start of each simulation, the secondary is at apastron, where
the star's initial influence on the system is minimal. We then run
each model twice, once with prograde and once with retrograde orbiting
particles. 

As the simulation progresses, we remove all particles beyond $2~\AU$
from the secondary; well beyond the Hill~radius. We then calculate the
eclipse duration at pericentre from the most distant particle on both
sides of the star--secondary axis ($d_{\rm min}$ and $d_{\rm max}$)
and the secondary's velocity relative to the star at that moment
($v_{\rm peri}$) (Equation~\ref{eq:eclipsetime}).

\begin{equation}
  \label{eq:eclipsetime}
  T_{\rm eclipse} = (d_{\rm max} - d_{\rm min}) / v_{\rm peri}
\end{equation}

\section{Size of the system}
\label{sec:size}
We simulate $10^5$ years ($\sim 9090$ orbits of the secondary) of
orbital evolution of the system, and compare the initial distribution
of particles to their final distribution, noting the radius within
which particles remain bound to the companion.

\begin{figure*}
  \centering
  \includegraphics[width=0.3\hsize]{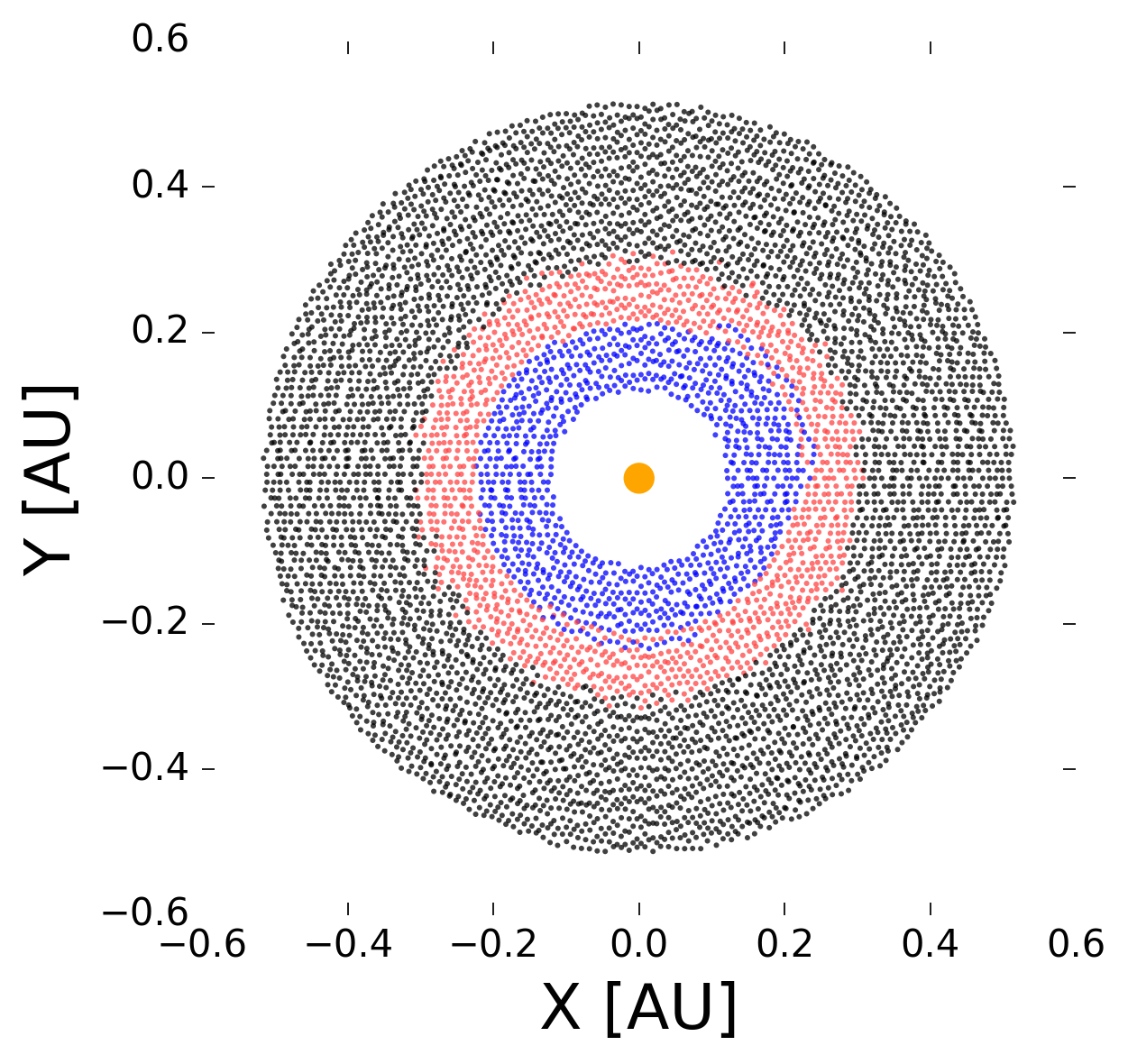}
  \includegraphics[width=0.3\hsize]{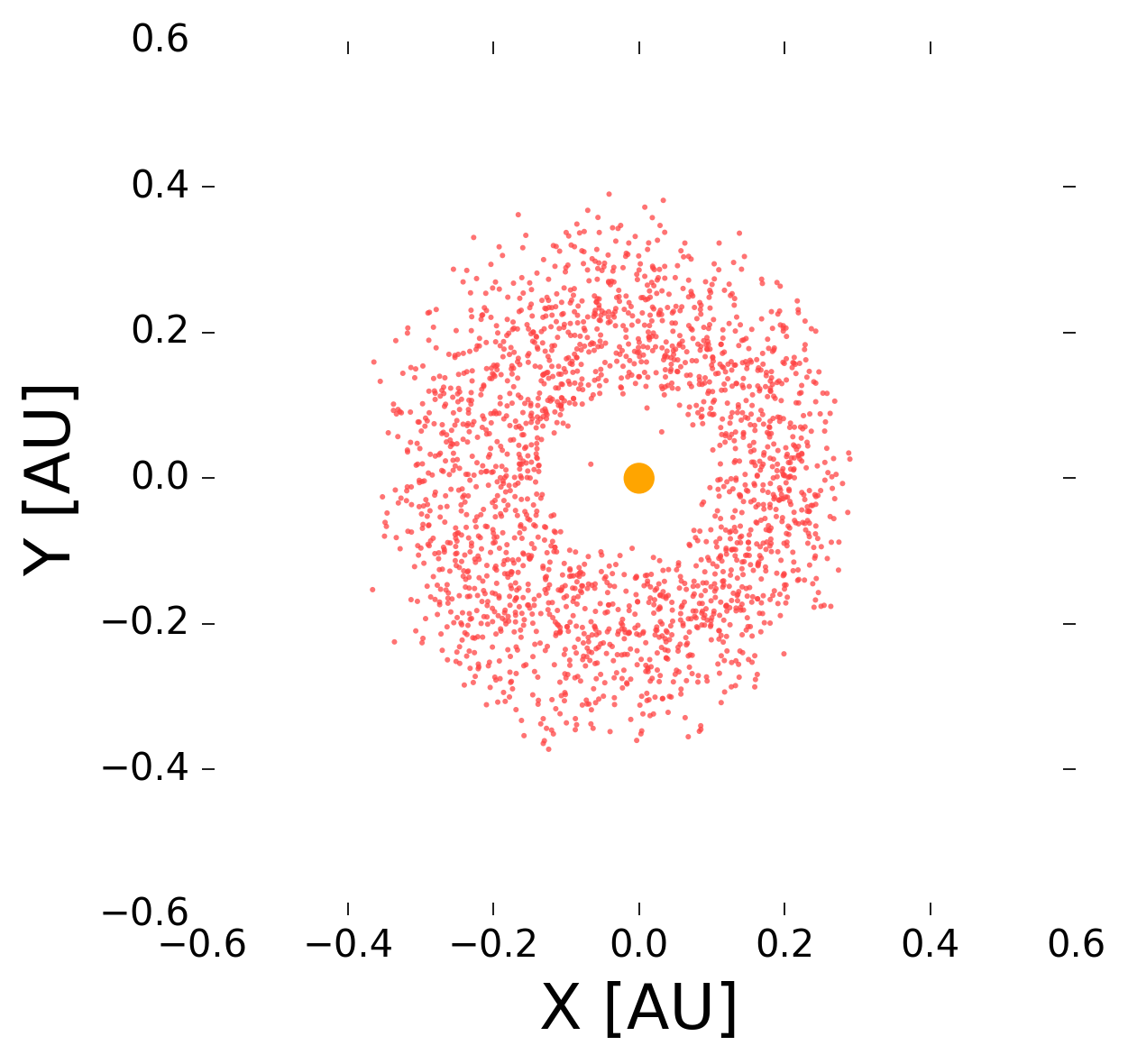}
  \includegraphics[width=0.3\hsize]{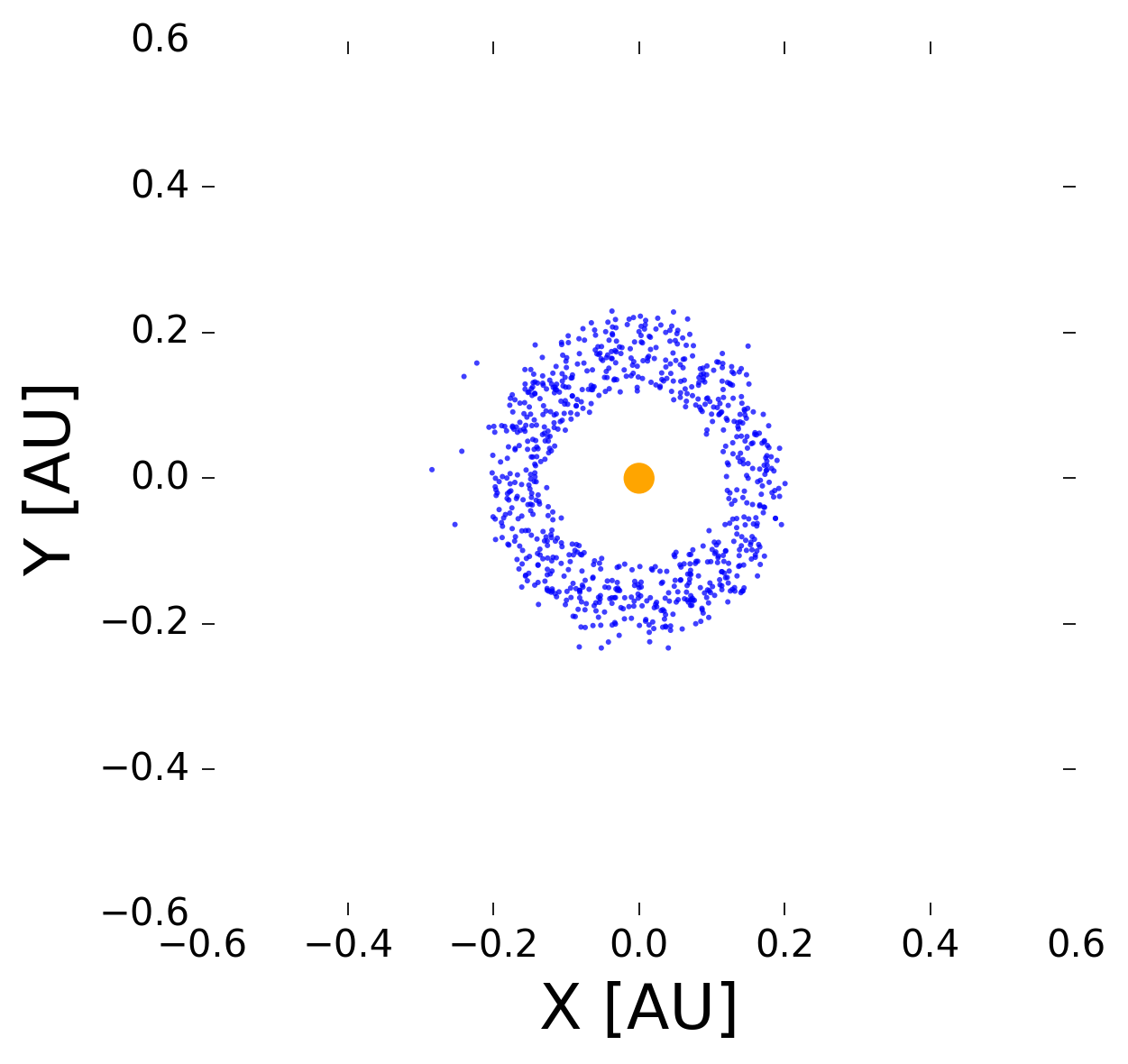}

  \caption{Particle distributions for model B80, initial configuration
    (left, at apocentre) and after $10^5$ years for the retrograde
    (middle) and prograde (right) cases, respectively; both at
    pericentre. In the initial configuration; black, red and blue
    particles represent particles that are stripped, bound in only the
    retrograde case and bound in the prograde case, respectively.
    \label{fig:ringconfiguration}} 
\end{figure*}

In Figure~\ref{fig:ringconfiguration}, we plot the particle density of
model B80, which has an initial radius of $0.53\AU$. In the left
panel, we show the system at initialisation; with black, red and blue
particles representing particles that are unbound, bound in the
retrograde case only and bound in the prograde case, respectively.
In the middle and right panels we illustrate the same system after
$10^5$ years of evolution for the retrograde and prograde case,
respectively, using red for the retrograde case and blue for the
prograde case. 

We find that prograde disc systems are severely disrupted, losing all
particles initially further out than $0.45\pm 0.01~\RHill$~within a
few passages; whereas retrograde systems are stable out to much larger
radii, gradually losing all particles with an initial radius over
$0.64\pm0.04~\RHill$.
This is in agreement with classical results from e.g.
\citet{ToomreToomre72}, who studied similar interactions on a galaxy
scale, and \citet{Morais12}, who studied 3-body systems on planetary
scale.

We also find that the initially circular ring particles become
increasingly eccentric due to the periodic forcing of the companion's
eccentric orbit and periastron passage (in a similar manner as the
effect described by \citet{Thebault06}).  This initially leads to
spiral patterns in the distribution of particles, which wind up, and
are difficult to detect after more than a few hundred orbits.

\begin{figure*}
  \centering
  \includegraphics[width=0.49\hsize]{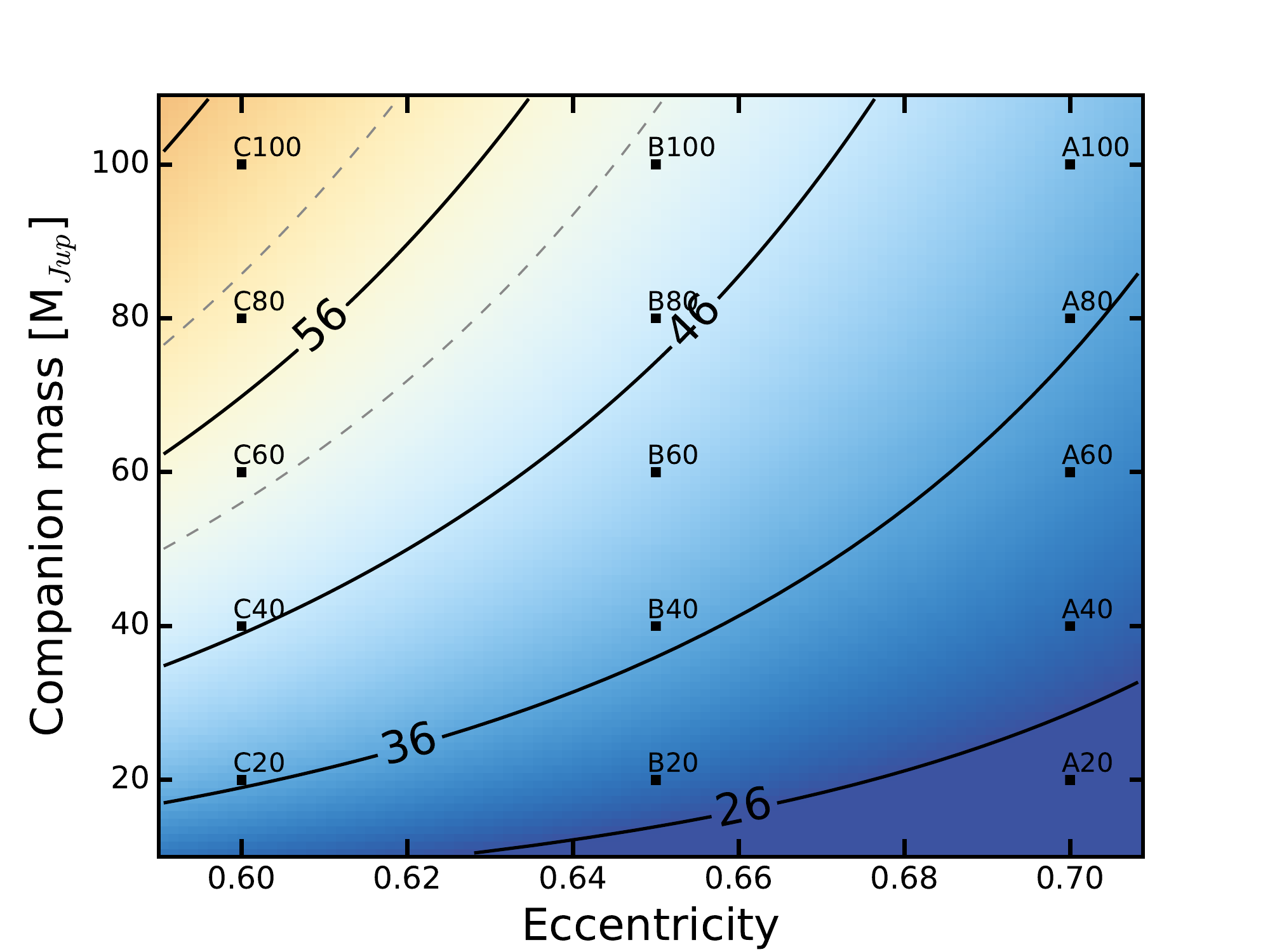}
  \includegraphics[width=0.49\hsize]{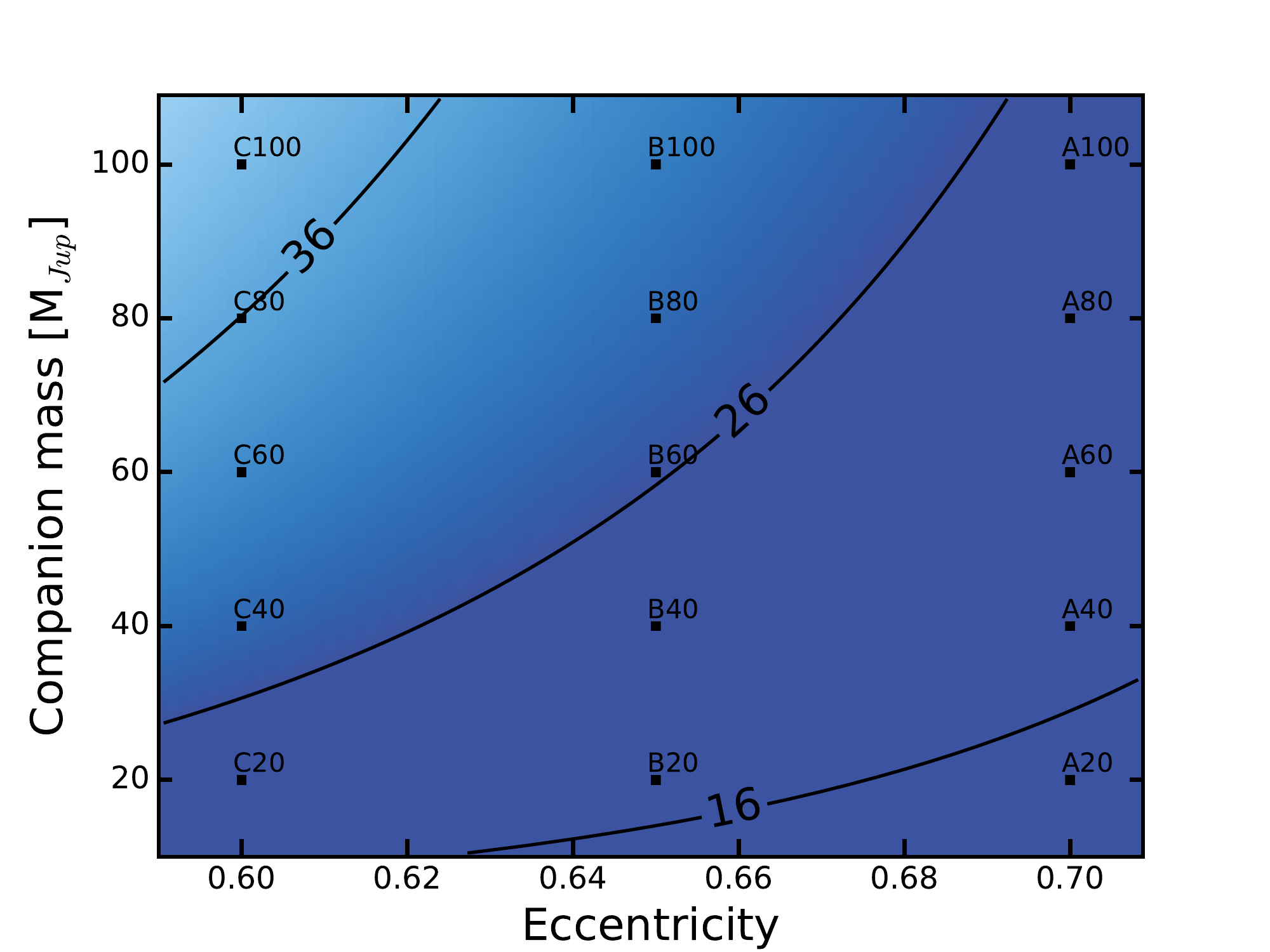}

  \caption{ Theoretical eclipse duration for retrograde (left) and
    prograde (right) systems, based on the size obtained from our
    simulations. Lines indicate equal eclipse durations (given in
    days), dashed lines are plotted for $56\pm4$~days. The positions
    of our models are indicated with squares.
    \label{fig:theoreticalduration}}
\end{figure*}

The radius after $10^5$~years, measured along the axis perpendicular
to the star--secondary axis, is $0.75\pm0.04 \times R_{\rm Hill}$ and
$0.46\pm0.02\times R_{\rm Hill}$ for retrograde and prograde systems,
respectively. We use these values to plot the theoretical eclipse
duration for all systems with an eccentricity between $0.59$ and
$0.71$ and a companion mass between $10$ and $110 \MJup$ in
Figure~\ref{fig:theoreticalduration}. 

\begin{figure}
\centering
\includegraphics[width=\hsize]{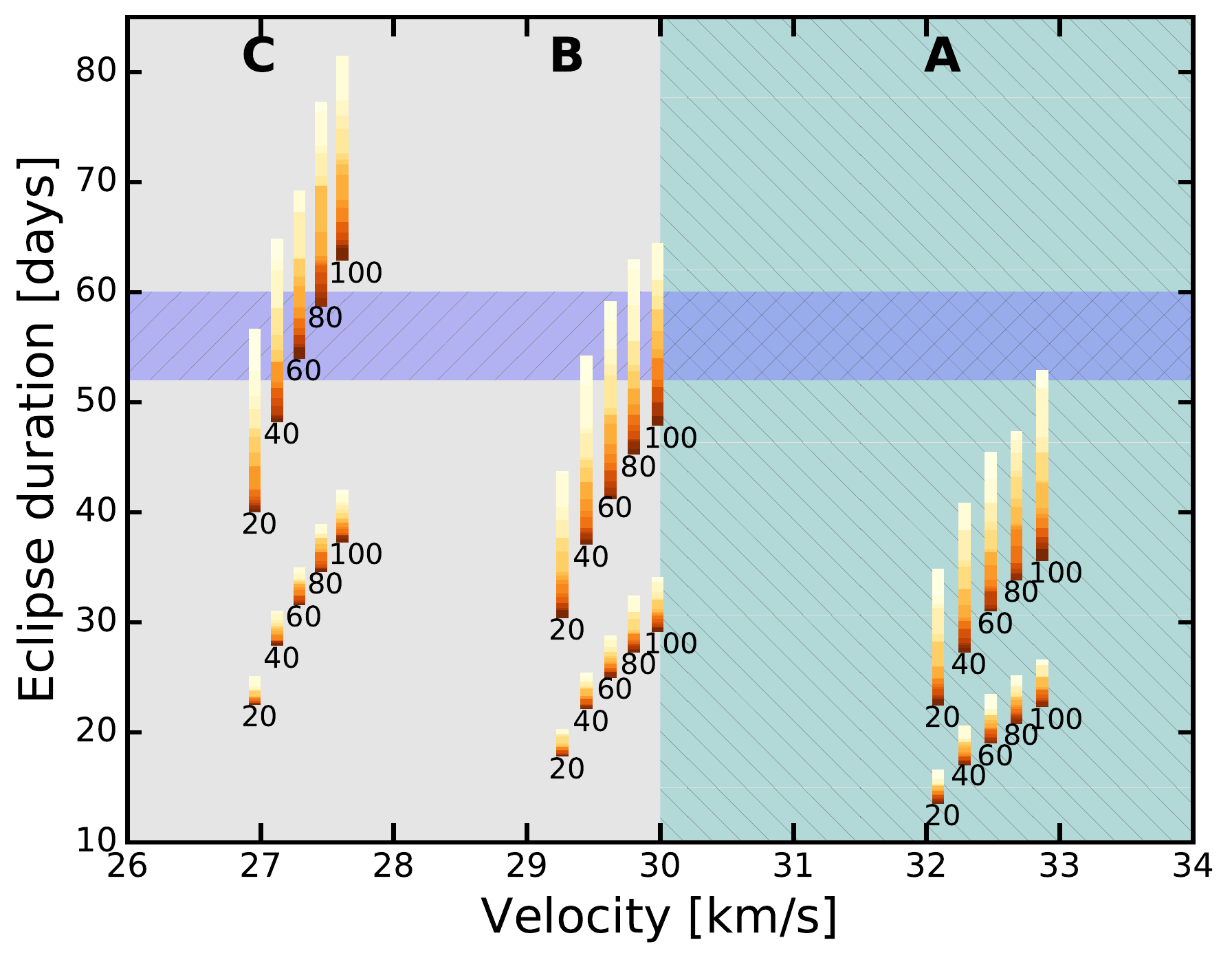}

  \caption{ Evolution of the eclipse duration at pericentre for the
    simulation models. The hashed blue region indicates the observed
    duration of $56 \pm 4$~days, while the hashed turquoise region
    indicates the observed velocity of $32 \pm 2~\kms$.
    The upper models with the longer eclipse duration are the
    retrograde models, the lower models are the prograde ones. The
    colour bar indicates the age of the model in ${\rm
    log}_{10}$-scale: lightest is after $3000$ years, while darkest is
    after $10^5$ years.  \label{fig:systemsize}}
\end{figure}

In Figure~\ref{fig:systemsize}, we show the evolution of the eclipse
duration for each of our models at pericentre between $3000$ and
$10^5$~years, combined with the expected values from the observations.
While the prograde systems only show a small change in eclipse
duration during this time, the change in the retrograde
systems is much larger, with the eclipse duration shortening by about
$10$~days. 

The eclipse durations are up to $\sim 40$~days for prograde models.
These models can therefore not explain the $56$~day duration of the
J1407b eclipse.  These models would require either a much more massive
companion or a much less eccentric orbit to work as a solution
(Equation~\ref{eqn:Hill}). Since a less eccentric orbit would result
in a lower transverse velocity, this would be inconsistent with the
derived velocity of $32 \pm 2~\kms$ from \citet{Kenworthy15}.
Similarly, \citet{Kenworthy15} restricts the mass of J1407b to be
$<100\MJup$ at $3\sigma$. As a result, any stable prograde orbit would
be inconsistent with the observed and derived orbital parameters of
J1407b.

In contrast, the eclipse lasts up to $\sim 80$~days for some of the
retrograde models. In particular the highest-mass B~models can have
eclipses that last long enough, while still being within the margin of
error for the derived velocity. If the system formed recently, lower
values for the companion's mass or orbits with higher eccentricity may
also provide a solution in agreement with the observations. We
therefore conclude that a giant ring system as proposed in
\citet{Mamajek12} can be a dynamically stable solution, but only if
the rings are on a retrograde orbit, for relatively large masses of
J1407b.

\section{Results and Discussion}
\label{sec:results}
\subsection{Results from the simulations}

Our main results from the simulations can be summarised as follows:

Retrograde ring systems can exist for at least $10^5$ years and
produce eclipses that last 56 days in duration with a projected
velocity consistent with that seen in the observational data (see
Figure~\ref{fig:systemsize}). 
No prograde ring systems are large enough to explain the duration of
the eclipse, since the outer rings are scattered within one or two
orbits of the companion.
Retrograde ring systems are typically $1.4$ times the size of prograde
ring systems, consistent with similar results from other simulations
\citep{ToomreToomre72,Morais12}.

We constrain the mass of the secondary companion to be larger than the
mass of 20 \MJup~expected by \citet{Kenworthy15}, closer to 100
\MJup~for the longest transit durations.

Initially circular particle orbits in the retrograde disc systems
become eccentric over time. As a result, particles starting within the
stable radius may be stripped. When the orbits change from circular to
more eccentric, spiral patterns are seen in the distribution of
particles, but these disappear over time.

\subsection{Discussion}

\subsubsection*{Lifetime of rings}
We simulate a disc of circumsecondary material for a period of
$10^5$~years. On this timescale, the discs shrink from $\RHill$ to
$0.46\RHill$ and $0.75\RHill$ for pro- and retrograde orbits,
respectively.  For prograde orbits, this process is quick (within
several orbits), whereas retrograde orbits shrink more gradually.

Since the system is significantly larger than its Roche limit (which
is on the order of $0.001\AU$ for a Moon-like object), satellites may
form in the disc. This is supported by the observed gaps in the
eclipse \citep{Kenworthy15b}. The timescale on which such satellites
would form may help constrain the age of the ring system around
J1407b.

Our simulations assume no new material is added to the system.
However, on a timescale of $10^5$~years, we consider it likely to be
replenished by material from a variety of sources, as J1407 is a young
and dynamically active system (see Section~\ref{sec:replenishment} for
a discussion of possible replenishment sources). The longer disruption
timescale of retrograde systems would allow them to keep their size
for a longer time in this case, while the disruption timescale for
prograde systems is too short.

\subsubsection*{Retrograde rings}

One of the strongest results out of our simulations is the implication
that the rings are in retrograde motion.
The additional stability of retrograde ring solutions over the
equivalent prograde ring solutions is to the point that we can rule
out all prograde ring solutions from being consistent with the
observations.
Explaining how a retrograde ring system in an highly eccentric orbit
around a young star came to be contributes additional complexity to
the giant exoring model presented in \citet{Kenworthy15}.

Two of the eight planets in our Solar system show evidence of giant
impacts that resulted in high planetary inclinations with respect to
the Ecliptic.
In these cases, both the rings and planet have a large enough obliquity
so that both planet and rings are retrograde.
It is possible that such a collision between two rocky bodies  in orbit
around a planet results in a significant amount of retrograde moving
material in the Hill sphere of J1407b, resulting in the rings we see
today.
Given that the height to diameter ratio of the J1407b exorings is $<
0.01$ \citep{Mamajek12} and that there is radial structure at all
spatial scales, the system is dynamically cold and any such retrograde
generating event must have happened a dynamically long time ago.

\subsubsection*{Replenishment of rings}
\label{sec:replenishment}

The ring lifetime may be boosted beyond that seen in our simulations
via a process of replenishment of the rings. For debris disks, a
collisional cascade from a reservoir of large bodies generates micron
sized material that is seen in reflected light observations
\citep{Backman93,Dent00}.

Processes that generate additional dust and ice in the J1407b Hill
sphere may well include the tidal disruption of bodies that are
captured by the gravity of J1407b, as suggested for the ice rings of
Saturn (Canup 2010) and as seen in our own solar system when comets
are tidally disrupted and captured by Jupiter, for example the tidal
disruption of the comet Shoemaker-Levy~9 in July 1992
\citep{Sekanina94} that resulted in a string of icy cores surrounded
by a dust cloud.

The presence of large numbers of comets in young planetary systems has
been seen towards the beta Pictoris system
\citep{Ferlet87,Lagrange-Henri88,Lagrange-Henri89,Kiefer14} and those
seen in transit towards KIC 8462852 \citep{Boyajian15}.

Two possible tests for this hypothesis are (i) to carry out
transmission spectroscopy of the ring system during the next eclipse
and determine the age of the rings by looking for recently ground up
material, and (ii) to see if the generated and disrupted dust is
observed at thermal and longer wavelengths and at radii larger than
the Hill sphere. 

\subsubsection*{Uncertainty dominated by radius of star}

The relative velocity of the star and ring system is derived from the
gradient of the light curve measured towards J1407 \citep[see Equation
11 in ][]{vanWerkhoven14}, $$v_{\rm ring} \propto \dot L
R_{*},$$ where the star is observed to change brightness at a
maximum rate of approximately $3L_{*} \rm{day}^{-1}$.  The diameter of
the ring system is simply $v_{\rm ring}T_{\rm eclipse}$ and relative
velocity of the ring system is directly proportional to the radius of
the star, and our greatest uncertainty in the size of the giant ring
system is dominated by the uncertainty of the stellar size
\citep{Kenworthy15}.

There is no direct measurement of the radius of the star, and it will
not be resolvable using ground-based interferometers in the forseeable
future, so we use the luminosity and effective temperature from
stellar evolution models to derive the radius of the star.
Additional constraints are set by rotational broadening of spectral
absorption lines measured using CORALIE \citep{Kenworthy15}, imposing a
radius of $R_* > 0.93\pm 0.02 R_\odot$.
If the assumed radius of the star is smaller than stellar models
predict, then the relative velocity and diameter of the ring system
become smaller. This also decreases the derived eccentricity of the
orbit of J1407b, easing the dynamical problem we face in this paper.

\subsubsection*{Non-azimuthal structures in the rings}

An alternative explanation is that there is additional internal
structure in the ring system that is non-azimuthally symmetric.
The vector addition of J1407b's orbital motion with the orbital motion
of ring material can explain the high relative velocity, discussed in
\citet{vanWerkhoven14}.
As shown by our simulations, non-radial structures develop during
periastron passage on dynamically short timescales that are comparable
to the duration of the transit.

\section{Conclusions}
\label{sec:conclusions}
We have performed simulations that consist of a companion on an
eccentric orbit consistent with the most probable orbital parameters
as detailed in \citet{Kenworthy15}.
A disk composed of particles initially orbiting the secondary in
circular orbits out to the Hill radius are added and the simulation is
run for $10^{5}$ years.
The particles are run in both a prograde orbital sense and a
retrograde sense.
As expected, the prograde ring system loses a significant portion of
its mass in a few orbits, and we do not find a stable prograde ring
system consistent with the observed orbital parameters and eclipse
duration.
For the retrograde ring system, we find it retains a larger fraction
of its mass out to larger radii, and for the proposed orbital
parameters of J1407b, a disk size and orbital velocity consistent with
observations is seen.

Circumplanetary disks are expected to be prograde with respect to the
circumstellar disks they formed in.
With a retrograde ring system, the question is raised as to how it
came into existence.
Uranus has a tilt of 98 degrees, with an associated ring system, but
no consensus of how it ended up with this obliquity.
Early theories suggested a single giant impactor caused the planet to
tilt over \citep{Safronov66}, and possibly disrupt an initially
circular orbit into the elliptical one we hypothesise.
More recently, \citet{Morbidelli12} show that a single impactor leads
to retrograde motions in the rings and moons, and that a series of
smaller impacts can preserve the orbital motion of the rings.

There is precedence for planets with retrograde orbits beyond our Solar
system.
Extrasolar planets have been detected with their orbital axes inclined
by more than 90 degrees with respect to their star's rotation axis
\citep[e.g. WASP-17b;][]{Anderson10}.
An interaction with a third companion in the system through the Kozai
mechanism \citep{Kozai62} is thought to provide the mechanism in these
cases.
The hypothesized third companion may still be within the J1407 system,
but a deep direct imaging search with Keck reveals no candidates within
400 AU greater than $6~\MJup$~\citep{Kenworthy15}.
An alternative explanation is that the third companion was ejected out
of the system and is a free floating object.
Evidence of strong gravitational scattering may be present in the
distribution of dust within the J1407 system, and observations at sub-mm
wavelengths with telescopes such as ALMA may provide additional
information.
One way to discriminate between these two hypotheses is to measure the
planet's obliquity and determine if it is greater than 90 degrees, but
this is not possible to test with current instruments.
Spectroscopic measurements during the next eclipse, however, can reveal
the orbital direction of the rings with respect to the rotation axis of
the star and confirm the retrograde ring hypothesis.

\begin{acknowledgements}
We are grateful to Nathan de Vries, Lucie J\'{i}lkov\'{a}, Masaki
Iwasawa, Erin~L. Ryan and Eric Mamajek for many useful discussions,
suggestions and comments that helped shape this work.  
We would also like to acknowledge the anonymous referee, whose
comments helped improve this article.
Finally, we are grateful to David Mikolas for alerting us to an
inconsistency in an earlier version of the article.
\end{acknowledgements}

\bibliographystyle{aa}
\bibliography{size_and_dynamics_of_j1407b}

\end{document}